\documentclass[
 ]{ceurart}

\usepackage{listings}
\usepackage{xcolor}%
\usepackage{xcolor}
\usepackage{tcolorbox}
\begin{document}

\copyrightyear{2022}
\copyrightclause{Copyright for this paper by its authors.
  Use permitted under Creative Commons License Attribution 4.0
  International (CC BY 4.0).}

\conference{Woodstock'22: Symposium on the irreproducible science,
  June 07--11, 2022, Woodstock, NY}

\title{AI-powered Code Review with LLMs: Early Results}

\tnotemark[1]
\tnotetext[1]{You can use this document as the template for preparing your
  publication. We recommend using the latest version of the ceurart style.}

\author[1]{Zeeshan Rasheed}[
email=zeeshan.rasheed@tuni.fi,
]
\cormark[1]
\fnmark[1]
\address[1]{Faculty of information technology and communication sciences),
Tampere University, Finland}

 \author[2]{Malik Abdul Sami}[%
 email=malik.sami@tuni.fi,
 ]
 \fnmark[1]

 \author[3]{Muhammad Waseem}[%
 email=muhammad.m.waseem@jyu.fi,
 ]
 \fnmark[1]
 \address[2]{Faculty of Information Technology, University of Jyväskylä, Finland}

\author[4]{Kai-Kristian Kemell}[%
 email=kai-kristian.kemell@helsinki.fi,
 ]
 \fnmark[1]
 \address[3]{Faculty of Mathematics and Natural Science, University of Helsinki, Finland}

 \author[5]{Xiaofeng Wang}[%
 email=xiaofeng.wang@unibz.it,
 ]
 \fnmark[1]
 \address[4]{Faculty of Engineering, Free University of Bozen Bolzano, Italy}

 \author[6]{Anh Nguyen}[%
 email=Anh.Nguyen.duc@usn.no,
 ]
 \fnmark[1]
 \address[5]{Department of Business and IT, University of South Eastern Norway}

 \author[7]{Kari Systä}[%
 email=kari.systa@tuni.fi,
 ]
 \fnmark[1]
 
 \author[8]{Pekka Abrahamsson}[%
 email=pekka.abrahamsson@tuni.fi,
 ]
 \fnmark[1]

\cortext[1]{Corresponding author.}
\fntext[1]{These authors contributed equally.}

\begin{abstract}
In this paper, we present a novel approach to improving software quality and development efficiency through Large Language Model (LLM)-based agents designed to autonomously review code, identify bugs and code smells, provide improvement suggestions, and optimize implementations. Unlike traditional static code analysis tools, our LLM-based agents have the capability to anticipate potential future risks in the code. This enables a dual benefit: enhancing overall code quality while supporting developer learning by promoting a deeper understanding of best practices and effective coding techniques. In future work, we aim to extend this approach toward an multi-agent based system capable of identifying a wide range of technical debt, including code, design, architecture, testing, documentation, build, and infrastructure debt.

\end{abstract}


\maketitle

\section{Introduction}

Large Language Models (LLMs) have emerged as a transformative force across various domains, offering unique capabilities in understanding, generating, and analyzing text \cite{treude2023navigating}, \cite{rasheed2024large}. These models, built on large datasets and advanced neural network architectures, demonstrate an ability to understand context and provide insights that was previously impossible \cite{radford2018improving, radford2019language, ouyang2022training, brown2020language}. The integration of LLMs into software development has led to significant advancements and intriguing possibilities, such as how code is written, reviewed, and optimized \cite{lu2023llama}. By utilizing LLMs, developers can tap into a deep well of coding knowledge and best practices, potentially elevating software quality to new heights \cite{fan2023large}, \cite{chen2021evaluating}, \cite{rasheed2024codepori}.

Despite the vast capabilities of LLMs, their application in the domain of code review and optimization remains underexplored \cite{xu2022systematic}. Code review is a critical phase in the software development lifecycle, aimed at identifying bugs, ensuring adherence to coding standards, and fostering knowledge sharing among developers \cite{li2022codereviewer}. Traditional code review processes and static analysis tools, often lack the depth to provide actionable feedback beyond the detection of syntax errors or known patterns of bugs \cite{tufano2022using}. This gap highlights a significant challenge: there is currently no LLM-based model specifically designed to enhance code reviews to identifying issues and suggesting optimizations and educating developers on best practices.

Addressing this challenge, our paper introduces a novel LLM-based AI agent system specifically tailored for the software development context. The proposed multi-agent system consists of four agents that identify code smells, potential bugs, and deviations from coding standards, and, crucially, provide actionable suggestions for improvement. These suggestions aim to optimize code and propose alternative approaches, thereby supporting a dual objective: enhancing code quality and promoting developer learning. Our multi-agent system represents a step forward from traditional static analysis tools by offering a proactive approach to code improvement and fostering deeper engagement with the principles of efficient and effective coding practices.

Looking ahead, our future research aims to evaluate the accuracy and efficiency of documentation updates generated by our agent-based system compared to manual methods. This will involve an empirical study in which developers test the proposed multi-agent system across projects from various domains. Additionally, we plan to extend this work toward a multi-agent system capable of identifying a wide range of technical debt, including code, design, architecture, testing, documentation, build, and infrastructure debt. Our contributions can be summarized as follows:

\begin{itemize}
    \item Developed a novel LLM-based multi agents system for enhanced code review, providing actionable improvements.
\end{itemize}

\begin{itemize}
    \item Distinguished our approach from traditional static analysis tools by focusing on proactive code optimization.
\end{itemize}

\begin{itemize}
    \item Proposed future research to extend the system’s capabilities and empirically validate its effectiveness.
\end{itemize}

The rest of the paper is organized as follows. We review related work in Section \ref{Related Work} and describe the study methodology in Section \ref{Methodology}. The initial results of this study are presented in Section \ref{Results}. We provide our future goal in Section \ref{Future Work} and the study is concluded with in Section \ref{Conclusions}.

\section{Related Work}
\label{Related Work}
Code review is an important part in the software development lifecycle and involves a significant amount of effort and time of reviewers \cite{bosu2013impact}, \cite{sadowski2018modern}. The focus among researchers on automating various aspects of the code review process is increasing, covering areas such as suggesting appropriate reviewers \cite{thongtanunam2015should}, \cite{zanjani2015automatically}, predicting locations for comments \cite{shi2019automatic}, \cite{hellendoorn2021towards}, recommending review comments \cite{gupta2018intelligent} and enhancing code quality \cite{tufano2021towards}.

Thongtanunam \textit{et al}. \cite{thongtanunam2015should} discovered that 30\% of code reviews face challenges with assigning the correct reviewers. To address this issue, they introduced RevFinder, a tool that recommends suitable code reviewers based on file locations. In response to the same challenge, Zanjani \textit{et al}. \cite{zanjani2015automatically} introduced cHRev, a platform that recommends reviewers for new code modifications by utilizing historical data from past code reviews. Their work concentrates on refining the initial phases of the code review process, whereas other scholars are committed to resolving the intricate difficulties that arise during various stages of code review.

Shi \textit{et al}. \cite{shi2019automatic} introduced the DACE framework, which combines CNN and LSTM technologies, to forecast if a section of code change will receive approval from reviewers. Hellendoorn \textit{et al}. \cite{hellendoorn2021towards} applied the Transformer architecture to address this challenge. Additionally, they explored the relationships between various code sections within a pull request by encoding each section and calculating attention scores among them to integrate the data. Li \textit{et al}. \cite{li2019deepreview} approached automatic code review from a multi-instance learning perspective, treating each code section as an instance with the goal of predicting the acceptance of a pull request.
Focusing on aspects linked to review comments, Siow \textit{et al}. \cite{siow2020core} suggest code reviews through a retrieval approach. They introduce CORE, an attentional model based on LSTM that aims to understand the semantic details in both the source code and its reviews by using multi-level embeddings. On a different note, Tufano \textit{et al}. \cite{tufano2021towards} employ deep learning strategies to automate another aspect of code review. They developed a Transformer-based model designed to modify contributors' code in order to meet the specifications outlined in review comments. 

Balachandran \cite{balachandran2013reducing} and Singh \textit{et al}. \cite{singh2017evaluating} advocate for the deployment of static analysis instruments to automatically identify violations of coding standards and prevalent errors. Regarding the automation of particular tasks in code review, the authors have put forward methods to enhance the allocation of reviewers.

Through an analysis of tools and methodologies that facilitate code review, Turzo \textit{et al}. \cite{turzo2024makes} determined that widely used code review platforms (such as Gerrit, Code Flow, Phabricator) generally provide similar fundamental features, with minimal automation support for tasks.
Concluding the review of related work, it is evident that there is a notable absence of a comprehensive tool based on LLMs that can conduct detailed code reviews and document best practices, in addition to detecting code smells, potential bugs, and violations of coding standards. To address this gap, we propose LLM based AI agent model designed for autonomous code review that not only identifies bugs but also offers suggestions and recommendations. Additionally, our model facilitates code optimization, further enhancing the quality and efficiency of the software development process.

\section{Research Method}
\label{Methodology}
This section outlines the methodology adopted in our study to explore the deployment and efficacy of a LLM-based AI agent within the software development, particularly focusing on the code review process. Our approach encompasses the design, development, and evaluation of an LLM-based AI agent system aimed at identifying potential issues in code and providing actionable recommendations. By dividing the methodology into specific components aligned with our research question, we ensure a structured and comprehensive examination of how LLM technology can be utilized to enhance software quality and development practices.

\begin{tcolorbox}[colback=green!2!white,colframe=black!75!black]
\textit{\textbf{RQ1.} How can a LLM-based AI agent effectively assist in code reviews by identifying potential issues and offering actionable recommendations?}
\end{tcolorbox}

This question emerged from the recognition of the limitations inherent in traditional code review processes and tools, which often fail to provide deep insights or actionable feedback for optimization and follow the best practices. Our research seeks to address this gap by exploring the potential of LLM-based agents to significantly improve the code review process, thus contributing to the development of higher-quality software. By utilizing the advanced capabilities of LLMs to understand context and provide suggestions, we aim to transform the code review process. This approach will enhances the efficiency and effectiveness of code reviews. Ultimately, our work aspires to establish a new standard in software engineering, where AI-driven reviews become essential in developing strong, innovative, and user-focused solutions.

\subsection{LLM-assisted Code Review (RQ1)}
The methodology for our proposed LLM-based AI agent system, designed to assist in code reviews, revolves around four specialized agents: the Code Review Agent, Bug Report Agent, Code Smell Agent, and Code Optimization Agent. Each agent is tasked with a distinct aspect of the code review process, utilizing LLM technology to analyze code repositories, identify issues, and suggest improvements.

\textbf{Code Review Agent}: The primary function of the Code Review Agent is to analyze source code and identify potential issues, including bugs, code smells, and deviations from established coding standards. This agent performs the initial assessment of the submitted code and then forwards its findings to subsequent agents for deeper analysis within the multi-agent workflow.

In this project, the agent was designed using prompt-based instructions, leveraging the GPT-4 model to perform code review tasks. All LLM-based agents communicate through a centralized coordination component, which provides a shared platform for inter-agent messaging and task delegation. The system connects to the LLM through an API key, and API requests are executed at each iteration of the multi-agent discussion process to enable collaborative analysis and code generation.

\textbf{Bug Report Agent}: This agent specializing in identifying potential bugs within the code, this agent analyzes patterns and anomalies that have historically been associated with software bugs.
This agent operates after the initial review phase and receives the preliminary findings from the Code Review Agent, enabling it to conduct a more targeted inspection. Using prompt-driven instructions and the GPT-4 model, the Bug Detection Agent performs focused diagnostic analysis and communicates its results through the centralized coordination component shared by all agents in the system. Each interaction is executed through API calls, ensuring smooth integration within the multi-agent pipeline.

\textbf{Code Smell Agent}: 
 
After receiving prior assessments from earlier agents in the workflow, the Code Smell Agent performs a more design-oriented evaluation of the code. Using prompt-based instructions and the GPT-4 model, it proposes refactoring strategies aimed at improving long-term code quality and reducing technical debt. All outputs are communicated through the system’s centralized coordination layer, ensuring seamless integration within the multi-agent process.

\textbf{Code Optimization Agent}:
The Code Optimization Agent is responsible for enhancing the efficiency, clarity, and overall performance of the submitted code. Building on the analyses conducted by earlier agents, this agent evaluates the structure and logic of the implementation to identify opportunities for improvement. It generates recommendations for refining algorithms, reducing redundancy, improving computational efficiency, and restructuring code where beneficial.

Using prompt-driven instructions and the GPT-4 model, the agent is capable of producing optimized versions of the code while preserving its intended functionality. Its outputs are transmitted through the system’s centralized coordination component, enabling smooth interaction with the other agents in the multi-agent workflow.

\section{Preliminary Result}
\label{Results} 
In this section, we present the study results of the proposed LLM-based multi-agent system for autonomous code review for bugs, code smells, and provide suggestions to optimize code. Our findings indicate that the proposed system demonstrated a strong capability in identifying a range of issues from minor bugs to significant code smells and inefficiencies across different programming languages and AI application domains. Below, we present the results of our LLM-based multi agent system in Section \ref{LLM based AI agent}, specifically reporting the outcomes of RQ1.

\subsection{LLM-Based AI Agent Results (RQ1)}
\label{LLM based AI agent}
The preliminary evaluation of the proposed multi-agent system shows that the agent based system is capable of autonomously reviewing source code and producing logical and actionable feedback across all agent roles. In initial tests, the Code Review Agent successfully identified structural issues and provided accurate summaries of potential risks in the submitted code. 

The Bug Detection Agent showed good performance in identifying logical inconsistencies, potential runtime errors, and weak or error-prone code structures. In several cases, it detected issues that traditional static analysis tools either missed or reported with very limited explanation. This suggests that LLM-based reasoning can complement existing rule-based approaches by providing context-aware insights supported by both semantic understanding and general programming knowledge.

The Code Smell Agent was effective in recognizing common anti-patterns and stylistic concerns that may lead to long-term maintainability challenges. Its refactoring suggestions were generally well aligned with accepted software engineering principles, often recommending improvements to naming conventions, function decomposition, and architectural clarity. These results support the notion that a language model can reliably detect design-level inefficiencies even with limited contextual information.

Finally, the Code Optimization Agent demonstrated the ability to propose modifications that improve code readability, reduce redundancy, and enhance performance. In several test cases, the agent generated optimized versions of code segments while preserving functionality, indicating practical utility for real-world development workflows. Across the system as a whole, the coordinated interaction between agents produced consistent and meaningful outcomes, illustrating the potential of multi-agent LLM architectures to support automated code review and technical debt reduction.





\section{Future Work}
\label{Future Work}
As we advance the integration of LLMs based agents into the code review process, our research take a step forward towards enhancing the efficiency and quality of software through AI-assisted code reviews. Looking ahead, our trajectory for future research encompasses several key areas aimed at further validating and expanding the capabilities of our LLM-based agent system.

A primary focus will be on evaluating the accuracy and efficiency of generated outcome of our proposed system in comparison to traditional manual methods. An empirical study will be conducted to analyze the developer discussion on code reviews for above mentioned projects, specifically targeting the identification of code smells and the documentation of bug reports. This study will extend to include a comprehensive examination of best practice documentation and insights from developer discussions.

The objective of this empirical research is twofold. First, to quantitatively and qualitatively assess the effectiveness of our proposed system in generating documentation that is accurate and more efficiently produced than manual efforts. Second, to explore the model's potential in contributing to a more streamlined software development process. By automating aspects of documentation and code review, we anticipate a reduction in the time developers spend on these tasks, allowing for a greater focus on core development activities.

Additionally, we aim to explore the educational impact of our system on software developers. By providing actionable feedback and suggestions for code improvement, there is potential for significant advancement in developer knowledge and adherence to best practices. Understanding the extent to which our model can contribute to developer education will be a key aspect of our future investigations.

Finally, we also aim to extend this approach toward a multi-agent system capable of identifying a wide range of technical debt, including code, design, architecture, testing, documentation, build, and infrastructure debt.

\section{Conclusions}
\label{Conclusions}
This paper introduced a novel approach specifically designed to enhance the code review process. By integrating four specialized agents, we showed a substantial improvement in identifying potential issues in code and providing actionable recommendations for optimization.

Our findings indicate that LLM-based AI agents can improve traditional code review processes, providing a dual benefit: enhancing code quality and facilitating developer education. The agents ability to detect a wide array of code smell, suggest meaningful optimizations, and promote best coding practices presents a notable advancement in the automation of software quality assurance. 

The implications of our research extend beyond immediate enhancements in code review efficiency and effectiveness. By showing the capability of LLM-based AI agents to improve software quality and developer knowledge, we open the door to future innovations in software development processes. Our planned future work, focusing on the comparative analysis of LLM-generated documentation against manual methods, aims to further explore the potential of our system to streamline software development practices.

\section{Acknowledgment}
We express our sincere gratitude to Business Finland for their generous support and funding of our project. Their commitment to fostering innovation and supporting research initiatives has been instrumental in the success of our work.

\bibliography{sample-ceur}

\end{document}